\documentclass[aps,pre,superscriptaddress,showpacs, reprint, nofootinbib]{revtex4-1} 

\usepackage{graphicx}
\usepackage{xcolor}
\usepackage{lipsum}
\usepackage{physics}

\usepackage{latexsym}
\usepackage{amssymb}
\usepackage{mathrsfs}
\usepackage{amsmath}

\usepackage{bm}
\usepackage{verbatim}
\usepackage{xcolor}
\usepackage{hyperref}
\usepackage{epsfig}

\begin{document}
\title{Inferring the flow properties of epithelial tissues from their geometry}
\author{Marko Popovi{\' c}} \thanks{Equal contribution.}
\affiliation{Institute of Physics, {\' E}cole Polytechnique F{\' e}d{\' e}rale de Lausanne, CH-1015 Lausanne, Switzerland}
\author{Valentin Druelle} \thanks{Equal contribution.}
\affiliation{Institute of Physics, {\' E}cole Polytechnique F{\' e}d{\' e}rale de Lausanne, CH-1015 Lausanne, Switzerland}
\affiliation{Biozentrum, University of Basel, Klingelberstrasse 70, 4056 Basel}
\author{Natalie A. Dye}
\affiliation{Max Planck Institute for Molecular Cell Biology and Genetics, Pfotenhauerstrasse 108, 10307 Dresden, Germany}
\affiliation{Cluster of Excellence Physics of Life, TU Dresden, 01307 Dresden, Germany}
\author{Frank J{\" u}licher}
\affiliation{Cluster of Excellence Physics of Life, TU Dresden, 01307 Dresden, Germany}
\affiliation{Max Planck Institute for Physics of Complex Systems, N{\" o}thnitzer Strasse 38, 01187 Dresden, Germany}
\author{Matthieu Wyart}\email{Contact: marko.popovic@epfl.ch, matthieu.wyart@epfl.ch}
\affiliation{Institute of Physics, {\' E}cole Polytechnique F{\' e}d{\' e}rale de Lausanne, CH-1015 Lausanne, Switzerland}

\begin{abstract}
Amorphous materials exhibit complex material proprteties with strongly
nonlinear behaviors. Below a yield stress they behave as plastic solids,
while they start to yield above a critical stress $\Sigma_c$. A key
quantity controlling plasticity which is, however, hard to measure is the
density $P(x)$ of weak spots, where $x$ is the additional stress
required for local plastic failure. In the thermodynamic limit $P(x)\sim
x^\theta$ is singular at  $x= 0$ in the solid phase below the yield
stress $\Sigma_c$. This singularity is related to the presence of system
spannig avalanches of  plastic events. Here we address the question if
the density of weak spots and the flow properties of a material can be
determined from the geometry of an amporphous structure alone. We show that a
vertex model for cell packings in tissues exhibits the phenomenology 
of plastic amorphous systems. As the yield
stress is approached from above, the strain rate vanishes and the
avalanches size $S$ and their duration $\tau$ diverge. We then show that
in general, in materials where the energy functional depend on topology,
the value $x$ is proportional to the length $L$ of a bond that vanishes
in a plastic event. For this class of models $P(x)$ is therefore
readily measurable from  geometry alone. Applying this approach to a
quantification of the cell packing geometry in the developing wing
epithelium of the fruit fly, we find that in this tissue $P(L)$ exhibits
a power law with exponents similar to those found numerically for a
vertex model in its solid phase. This suggests that this tissue exhibits
plasticity and non-linear material properties that emerge from
collective cell behaviors and that these material properties govern
developmental processes. Our approach based on the relation between
topology and energetics suggests a new route to outstanding questions
associated with the yielding transition.
\end{abstract}

\maketitle

\section*{Introduction}

A fascinating aspect of biological systems is their ability to grow into well-defined shapes \cite{Thompson45}.
Thinking about tissues as materials, what should their properties be to allow for robust morphogenesis? 
One view is that tissues are viscoelastic fluids, molded into desired shapes by surface tension and active forces \cite{Bittig2008, Ranft2010, Lee2011, Blanch-Mercader2014, Etournay2015, Banerjee2015, Popovic2017, Julicher2018}. An alternative picture is that they are yield stress materials \cite{Mongera2018} similar to clay. Such materials allow for great control, since shape is changed only if the magnitude of shear stress $\Sigma$ is above the threshold yield stress $\Sigma_c$. These approaches can be thought of as two extremes of a continuous spectrum of models, since at finite temperature, or at finite level of active stress and cell divisions in biological systems \cite{Ranft2010, Matoz-Fernandez2017}, materials always eventually flow. Experimental evidence of glassy behaviour \cite{Angelini2011, Nnetu2012, Schotz2013}  indeed suggests relevance of intermediate cases. Quantitatively, an interesting observable  to distinguish these regimes is the ratio between the strain rate increment $\delta \dot \gamma$ and the  stress increment  $\delta \Sigma$ causing it. This ratio is simply $\dot\gamma/\Sigma$ for a Newtonian liquid, but  is infinite at $\Sigma_c$ in a yield stress material at zero temperature. As discussed below, this divergence is associated with  collective events where large chunks of the material rearrange. This fact suggests that one may be able to decide in which regime tissues operates simply by imaging their dynamics and geometry. One of our aims is to build the first steps of this long term goal. Note that this endeavour is distinct from non-invasive force inference methods \cite{Ishihara2012, Chiou2012}, in which one seeks to reconstruct stress - instead of plasticity and rheological properties - from geometry.

As it turns out, there is currently a considerable interest in understanding the relationship between geometry and plasticity in particulate amorphous materials \cite{Cubuk15,Gartner16,Patinet16,Wijtmans17,Schwartzman-Nowik2019}.  Flow is mediated by local rearrangements termed shear transformations \cite{Argon79} that are coupled by long range elastic interactions \cite{Picard04}. If thermal fluctuations are small, for $\Sigma>\Sigma_c$ the flow consists of avalanches of correlated shear transformations. The characteristic avalanche size diverges at $\Sigma_c$ \cite{Lemaitre09,Nicolas18}, and is system spanning for non-stationary slow (quasi-static) flows occurring in the solid phase $\Sigma<\Sigma_c$ \cite{Lin15}. At a macroscopic level, for $\Sigma \geq \Sigma_c$ the shear rate is singular and follows the Herschel-Bulkley   law $\dot\gamma\sim (\Sigma-\Sigma_c)^\beta$ \cite{Herschel1926}. A key ingredient of the scaling theory of this phase transition  \cite{Lin14} is that the density $P(x)$ of shear transformations at a distance $x$ to their local yield stress (i.e.  the shear transformations that yield if the shear stress is increased by $x$) follows $P(x)\sim x^\theta$, with $\theta > 0$ \cite{Lemaitre07,Karmakar10a,Lin14a}, which is directly related to the presence of extended avalanches \cite{Muller14}. The exponent $\theta$ is predicted \cite{Lin16} to vary non-monotonically as shear strain is increased  from an isotropic state,
as observed in particle-based models \cite{Wencheng19,Ozawa18,Shang19}, whereas for $\Sigma>\Sigma_c$ there are no singularities in $P(x)$ and $\theta=0$ \cite{Lin16}. Various computationally expensive numerical methods are being developed to extract  the field of shear transformations and their associated distance to yield stress $x$  from the structure alone \cite{Cubuk15,Gartner16,Patinet16,Wijtmans17,Schwartzman-Nowik2019}, as it would allow one to study fundamental questions including the possible localization of plastic strain. In this work we present an alternative approach by showing that some models of disordered materials present the same phase transition, in which this extraction is straightforward. 

We consider the vertex model of epithelial tissues \cite{Farhadifar2007} in its solid phase where it displays a finite elastic modulus \cite{Bi2015, Bi2016}. We first show that its yielding transition is similar to particulate amorphous materials: its flow curve is singular with a Herschel-Bulkley exponent $\beta\approx 1.3$, associated with  avalanches of plastic events whose  size  diverges  as $S\sim \dot\gamma^{-a}$ and last a duration $\tau\sim \dot\gamma^{-c}$ where  $a\approx 1/4$ and $c \approx 2/3$. In such a model, just like for dry foams, shear transformations are known to correspond to T1 events \cite{Kabla03}. We argue quite generally and test numerically that in models where the energy function depends on the topology, the distance $x$ to  the local yield stress follows $x\sim L$ where $L$ is the bond length between two vertices. It implies that $P(x)$ is readily obtainable from the bond length distribution, from which we extract $\theta$. It is found to change non-monotonically under strain
with $\theta\approx 0.5 - 0.6$ in isotropic state and with a similar value at large strain at $\Sigma=\Sigma_c$: $\theta\approx 0.56-0.77$. By contrast, $\theta$ vanishes in the liquid phase for $\Sigma>\Sigma_c$. Finally, we measure the bond length distribution in fruit fly wing disc and pupal wing epithelia, and find a similar behaviour with $\theta\approx 0.7-0.9$. This measurement suggests the existence of collective effects in tissues 
and raises the possibility that these materials function in a regime of high sensitivity $\delta \dot \gamma/\delta \Sigma \gg \dot\gamma/\Sigma$.

\section*{Flow and loading curves of vertex model}

We use the standard vertex model of epithelial tissues \cite{Farhadifar2007, Staple2010} where the 2D network of polygonal cells is assigned an energy function
\begin{align}
\label{eq:4}
E= \sum\limits_{c \in \text{cells}}\frac{1}{2}\left[K \left( A_c - A_{0, c} \right)^2 +\Gamma_c P_c^2\right]+ \sum\limits_{b \in \text{bonds}}\Lambda_b L_b   \quad ,
\end{align}
where $A_c$, $L_b$ and $P_b$ are cell area, bond length and cell perimeter, respectively.
Model parameters specify preferred cell area $A_{0, c}$, bond tensions $\Lambda_b$ and cell perimeter stiffness $\Gamma_c$.  To avoid localisation of flow in a narrow shear band, which occurs in the homogeneous vertex model \cite{Merkel2014}, we introduce cell size polydispersity for shear flows, see Supplementary Information (SI). In all simulations the network is in solid phase with the normalised preferred perimeter $p_0\equiv - \Lambda_b/(2 \Gamma_c \sqrt{A_{0,c}})\simeq 3.41$ for all cells, well below the rigidity transition point $p_0^{*}\simeq 3.81$ \cite{Bi2015}. Note that our results below may not hold in the fluid phase of the vertex model \cite{Park2015}, or in active tension networks with isogonal soft modes \cite{Noll2017}.

The dynamics of the cellular network is described by overdamped dynamics of vertex positions $\vec{r}_{\alpha}$
\begin{align}
  \frac{d\vec{r}_{\alpha}}{dt}= - \nu \vec{\nabla}_{\alpha} E \quad ,
  \end{align}
  where $\nu= 1$ is a mobility. A T1 transition occurs when a bond length becomes smaller than a threshold length $\epsilon_{T_1}$, see SI.

\begin{figure}[ht!]
\centering
\includegraphics[width=.48\textwidth]{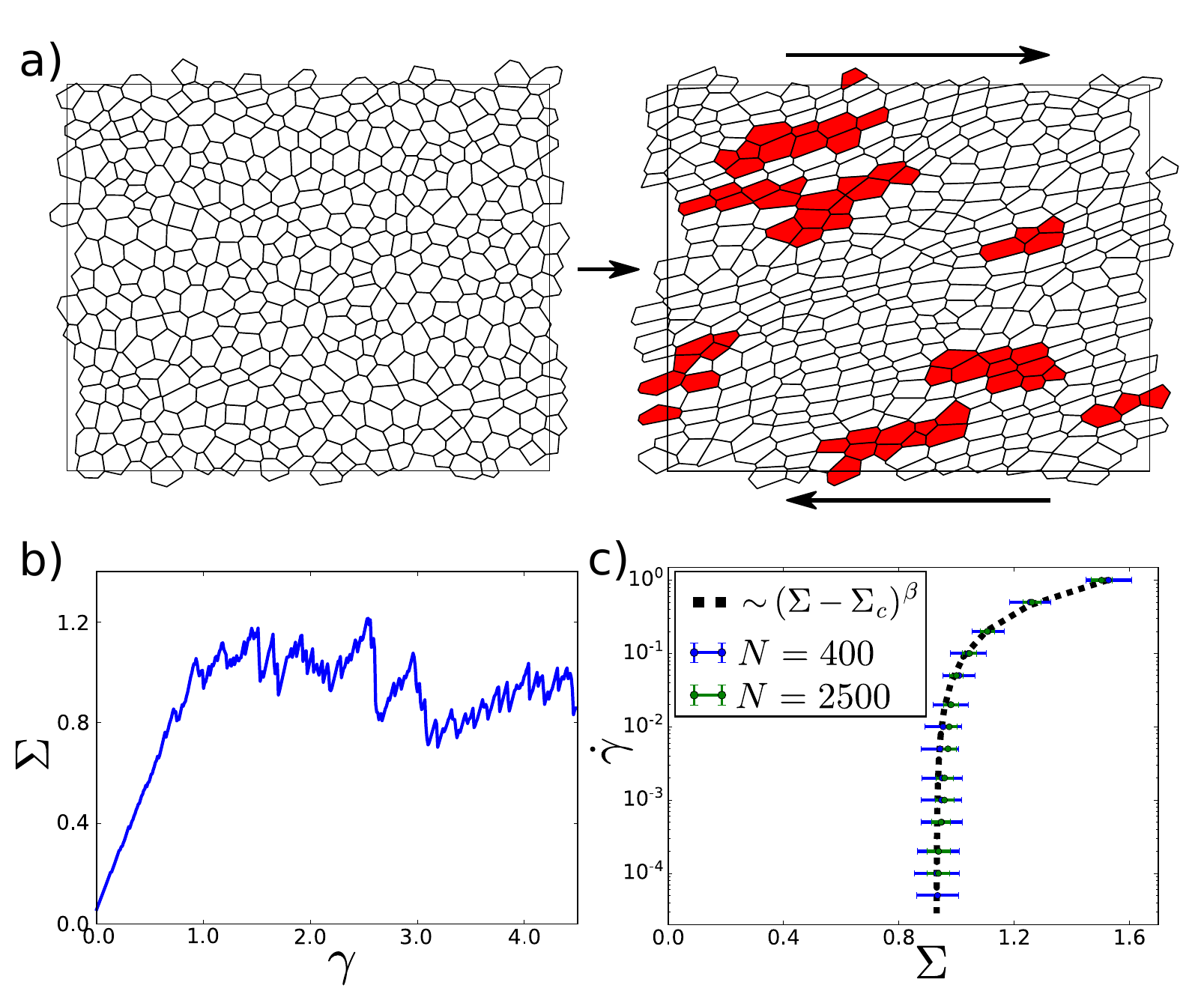}
\caption{\textbf{a)} Left: Isotropic disordered polydisperse periodic network of $N= 400$ cells. Right: Snapshot of a network in a steady state simple shear flow. Colored cells recently participated in a T1 transition. \textbf{b)}
Initial elastic regime of network under shear strain is followed by plastic regime where elastic loading and relaxation by T1 transitions are equilibrated. 
Ruggedness of the stress \textit{vs} strain curve appears due to periods of stress increase through elastic loading and stress drops by T1 rearrangements. \textbf{c)} Steady state flow in two different network size: 400 cells in blue and 2500 cells in green. Dotted line shows the fit of Herschel-Bulkley law.}
\label{fig:rheology} 
\end{figure}

We use two ensembles of isotropic disordered networks with $N= 400$ and $N= 2500$ cells, as described in SI. An example of a network with $N=400$ cells is shown in Fig. \ref{fig:rheology} a) left. We perform simple shear strain simulations on these networks at a constant strain rate, illustrated in \ref{fig:rheology} a) right for $\dot{\gamma}=10^{-4}$. The network initially responds elastically: the shear stress $\Sigma$ in the network grows almost linearly (Fig. \ref{fig:rheology} b)). As the strain is increased, T1 transitions occur and relax the stress in the network, visible as sudden drops in the stress \textit{vs} strain curve. In Fig. \ref{fig:rheology} a) right we visualise recent T1 transitions that occurred during a strain increment $\Delta \gamma= 0.15$ by coloring participating cells in red. T1 transitions appear to be correlated and organised into avalanches of various sizes, corresponding to widely distributed stress drops in Fig. \ref{fig:rheology} c).  Eventually a steady state is reached in which stress relaxation due to T1 transitions balances the elastic loading. In Fig. \ref{fig:rheology} d we show the steady state flow curve. It is  well described by the Herschel-Bulkley law $\dot\gamma\sim (\Sigma-\Sigma_c)^\beta$ \cite{Herschel1926} with the yield stress $\Sigma_c\approx 0.93$\footnote{The stress component corresponding to the simple shear is defined by $\Sigma\equiv (\partial E/\partial \gamma) / N $. In our simulations $\Gamma = 1$ and a typical bond length is $L_0= 1$. Therefore, the reported values of stress can be understood as normalised by $\Gamma L_0^2$.} and exponent $\beta \approx 1.3$\footnote{A precise measurement of the exponent $\beta$ that could discriminate theoretical predictions \cite{Ferrero2019} is beyond the scope of this work.}. 

\section*{Energy cusp at T1 transitions}
\begin{figure}[ht!]
\centering
\includegraphics[width=.48\textwidth]{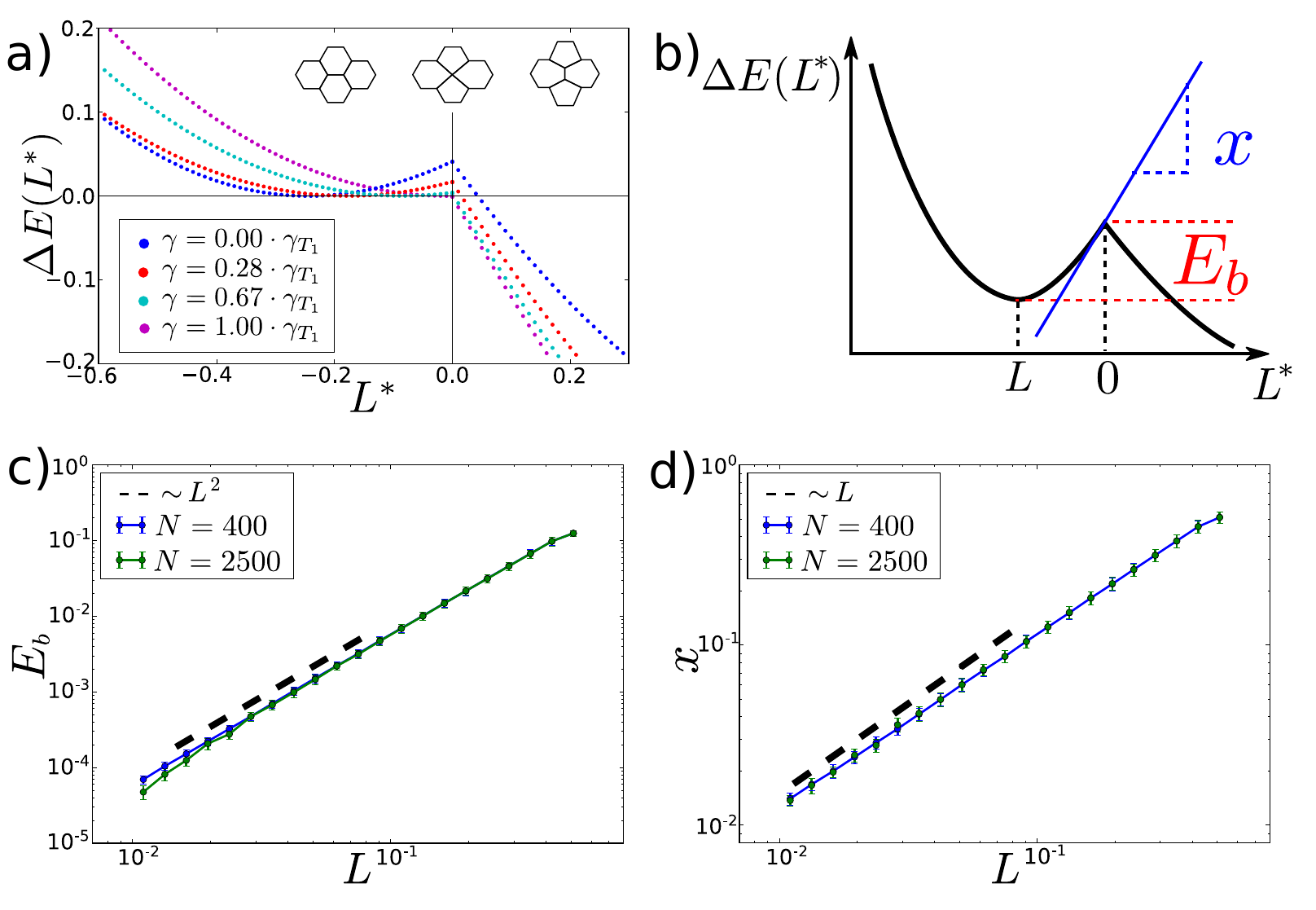}
\caption{\textbf{a)} The energy profile parametrised by the imposed length of a bond $L^{*}$ evolves under strain until the metastable state disappears at strain $\gamma_{T_1}$ and a T1 transition occurs. Negative values of bond lengths denote lengths before the T1 transition. \textbf{b)} Schematic of the cusp in the energy profile: the energy barrier is quadratic in bond length $E_b \sim L^2$ and the force required to shrink the bond  is linear $x\sim L$.  Test of scaling relations between the length $L$ and: \textbf{c)} the energy barrier $E_b$, \textbf{d)}  the force distance to a T1 transition $x$,  in isotropic disordered networks. Error bars represent one standard deviation of the sample.}
\label{fig:T1mechanics} 
\end{figure}

Next we characterize the T1 transitions or elementary plastic events.
For this purpose, we identify the bond in a network that will first disappear under strain. We then constrain the length of that bond to a value $L^{*}$ and determine the energy of the network under strain. The original bond length is assigned negative values and the new bond that appears through the T1 transition is assigned positive values. In Fig. \ref{fig:T1mechanics} we show the energy of the network relative to the energy of the unconstrained network, as a function of strain $\Delta E(L^{*}; \gamma) \equiv E(L^{*}; \gamma) - E(\gamma)$. Originally the system is in a metastable state, corresponding to the local minimum of $\Delta E$. As the shear stress increases the minimum disappears and the T1 transition occurs. The energy profile shows a cusp at the onset of T1 where $L^{*}= 0$, a well known feature of the vertex model energy landscape \cite{Bi2014, Bi2015, Su2016, Krajnc2018}. The presence of a cusp in the energy profile allows us to relate the bond length of short bonds  to the additional force\footnote{Note that exerting a stress increment at the boundary of the system will in general generate a bond force proportional to that increment, so the quantity $x$ we use here characterizes well the  distance to a local yield stress.} $x$ at the bond needed to drive a T1 transition.  Namely, expanding $\Delta E(L^{*}, \gamma)$ in bond length around the equilibrium value $L$ reads:
\begin{align}
  \label{eq:1}
  \Delta E(L^{*})\approx \Delta E(L) + \frac{1}{2}\Delta E''(L)(L^{*} - L)^2
  \end{align}
  and we see that at T1 transition, corresponding to $L^{*}= 0$, the energy barrier to the T1 transition is:
\begin{align}
  \label{eq:2}
  E_b \equiv \Delta E(L^{*}= 0) \approx \frac{1}{2}\Delta E''(L) L^2
  \end{align}
Thus, the force on that bond required to trigger the T1 transition is:
\begin{align}
\label{eq:3}
x&\approx \Delta E''(L) L \quad .
\end{align}
Since the effective stiffness of the bond $\Delta E''(L)$ is expected to be finite at a T1 transition, we find $E_b \sim L^2$ and $x \sim L$. This relationship between geometry and plasticity follows from presence of the cusp in the energy profile. Ultimately, the origin of this cusp lies in the form of the energy function that depends on the cell perimeters and area. These quantities are smooth functions of the vertex positions for a given network topology. However they are not smooth, but simply continuous, at the point where two vertices meet and the network topology changes. Consequently, forces can change discontinuously at the transition point. Therefore, we expect to generically find $x \sim L$ in cellular systems such as epithelial tissues and dry foams for which the dependence of the energy on the vertex position is topology-dependent. By contrast, particle systems in which the energy depends on the particle positions independently of any notion of topology cannot show such a cusp (as long as the interaction potential is smooth). Furthermore, due to the cusp at the T1 transition the stiffness of the corresponding displacement mode does not vanish, as it would at the plastic event in particle systems\footnote{In systems with smooth energy function a plastic event corresponds to the usual saddle-node bifurcation.}. Therefore, we do not expect to find a signature of local plastic events in the eigenvalues of Hessian matrix of energy function, which could explain the lack of soft non-localised modes recently observed in the Voronoi vertex model\footnote{The voronoi  vertex model has the same energy function as the usual vertex model, but its degrees of freedom correspond to cell centers. The network topology  is constructed from these centres by performing a Voronoi tessellation at each time-point.} \cite{Sussman2018}.

We test these predictions in isotropic disordered networks by forcing bond length to attain a very small value $L_{\text{min}}= 10^{-6}$ and determining the corresponding network energy change and the constraining force magnitude $x$. Results shown in Fig. \ref{fig:T1mechanics} c) and d) are consistent with our predictions. Therefore, identifying the locations of short bonds allows us to read the map of ``weak spots'' in the system, as well as to deduce the distribution $P(x)$.

\begin{figure}[ht!]
\centering
\includegraphics[width=.48\textwidth]{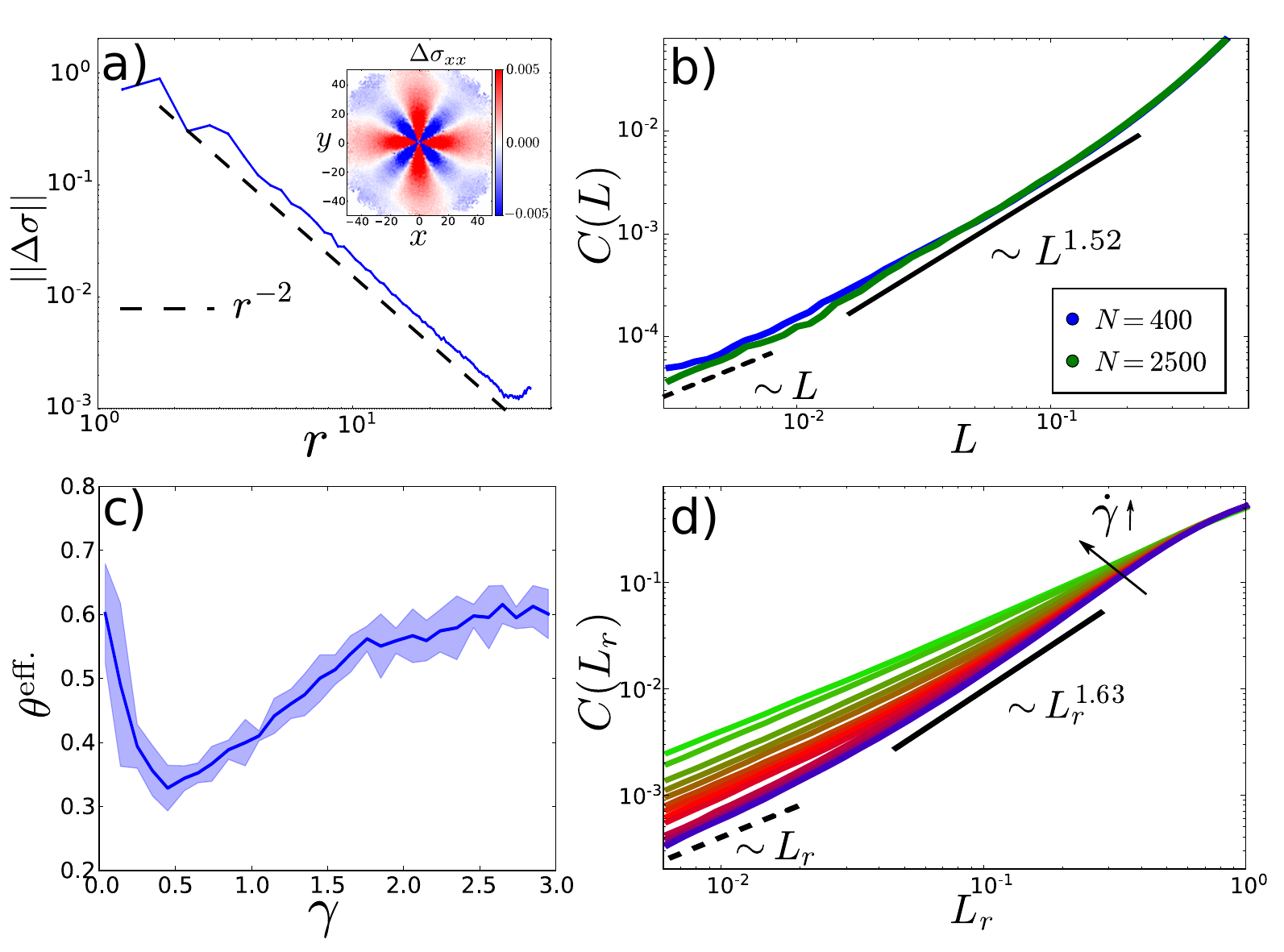}
\caption{\textbf{a)}: Magnitude of the shear stress redistribution $\norm{\Delta \sigma}$ in the cellular network after a  T1 transition at the origin. It is consistent with that of a force dipole. Inset: Four-fold symmetry of shear stress redistribution component $\Delta\sigma_{xx}$, consistent with that of a force dipole.  \textbf{b)} Cumulative bond length distribution $C(L)= \int_0^LP(L')dL'$ in disordered isotropic networks. It is consistent with $P(L) \sim x^{\theta}$ with $\theta \approx 0.5- 0.6$, see SI for details. At low $L$, $P(L)\sim \text{const.}$ due to finite system size. 
 \textbf{c)} The effective exponent $\theta^{\text{eff.}}$ measured as a function of strain, starting from a disordered isotropic networks, see SI.  
 \textbf{d)} The cumulative distribution of adjusted bond lengths $L_r \equiv L - \epsilon_{T1}$ in steady state simple shear flow with strain rate varying between $\dot{\gamma}= 5\cdot 10^{-5}$ (blue line) and $\dot{\gamma} = 1$ (bright green line). We find that at the lowest strain rates the effective exponent $\theta^{\text{eff.}}$ converges to  $\theta\approx 0.56 - 0.77$, see SI for details. As in isotropic networks, at low $L_r$ we find
 $P(L_r) \sim \text{const.}$, as expected for any finite strain rate \cite{Lin16}.}
\label{fig:stability} 
\end{figure}

\section*{Stability of the cellular network}
After a T1 transition, the network relaxes to a new metastable state with redistributed shear stresses. We measure the stress redistribution by enforcing a T1 transition and measuring the change in stress within each cell after the network had relaxed (the cellular stress is defined as in \cite{Aliee2013}). In an elastic 2D medium we expect the shear stress redistribution to be consistent with that of a force dipole in an elastic medium: $\Delta\sigma_{xx}\sim \cos{(4\varphi)}/r^2$, $\Delta\sigma_{xy}\sim \sin{(4\varphi)}/r^2$ \cite{Picard04}.  
Fig. \ref{fig:stability} a) shows the shear stress redistribution, obtained by orienting the disappearing bond direction along the $x$-axis and averaging over 50 realisations, as detailed SI. We find a clear four-fold symmetry of $\Delta\sigma_{xx}$ (inset)  as well as inverse quadratic decay of its magnitude, as expected for a force dipole and consistent with simulations of 2D foams \cite{Kabla03}.

The stress change after a T1 transition can trigger new T1 transitions if there are bonds with small $x$ in the network. In the solid phase, the stability of the network with respect to extensive avalanches of T1 transition imposes $P(x) \sim x^{\theta}$ with $\theta > 0$, otherwise never-ending avalanches would occur \cite{Lin14a}. As we have demonstrated $x \sim L$ in the vertex model. Therefore, bond length distribution should vanish with the same exponent $P(L) \sim L^{\theta}$,  and $\theta$ can be extracted from $P(L)$. 
We measure the cumulative distribution $C(L) \equiv \int_0^L P(L')dL'$ in disordered isotropic networks. We find a scaling regime,  whose range of validity grows with system size, for which  $\theta\approx 0.5- 0.6$ (Fig \ref{fig:stability} b), see SI for details. At even smaller $L$, the bond lengths distribution departs from this scaling as $P(L) \sim \text{const.}$, as expected due to finite size effects and also observed in elasto-plastic models \cite{Lin14}.

Interestingly, these values are consistent with those found in two-dimensional elasto-plastic models \cite{Lin14}. In these coarse-grained models, the material is described as a collection of mesoscopic blocks with a simplified description of plastic events: a block yields when the local yield stress is reached, it accumulates plastic strain and redistributes stress in the material as a force dipole \cite{Baret02, Picard05}. Since both ingredients are present in vertex model as well, as we have seen,  it is not surprising that we find a consistent value of $\theta$.

We next studied the evolution of the exponent $\theta$ during the transient loading period, between the initially isotropic network and the steady state. Surprisingly, it has been predicted that $\theta$ would then non-monotonically depend on strain \cite{Lin16}, a result observed in elasto-plastic models \cite{Lin15} but only indirectly observable in amorphous solids where $P(x)$ is very hard to access \cite{Ozawa18,Shang19, Wencheng19}.
To test directly this prediction, we measure the bond length distribution as a function of strain at a small constant strain rate $\dot{\gamma}= 10^{-4}$, close to the quasi-static limit. Note that even in the thermodynamic limit we expect to find singular $P(x)$ and $P(L)$ only in the quasi-static limit of vanishing strain rate (at any finite rate, there are always T1 transition occurring leading to $\theta=0$ \cite{Lin16}. However, in a finite system we can still measure an effective exponent $\theta^{\text{eff.}}$. 
We confirm that the evolution of $\theta^{\text{eff.}}$ with strain is non-monotonic, see Fig. \ref{fig:stability} c).

It is also important to quantify the distribution of bond lengths in steady state flow, see Fig \ref{fig:stability} d). At high strain rates we find $\theta^{\text{eff.}} \to 0$ as expected, while in the limit of vanishing strain rates the effective exponent approaches the value $\theta\approx 0.56 - 0.77$ (Fig. \ref{fig:stability} d)). Thus, $P(L)$ can be used to locate  the distance to the yield stress, at least in this  setting where noise is absent. 

\section*{Collective behaviour of T1 transitions}

Quite generally in disordered systems \cite{Muller14}, a singularity in the density of weak regions $P(x)$ is synonymous to avalanche-type response
where many weak regions - here T1's - rearrange in concert. To test if this idea holds in the vertex model,
we quantify the correlation between T1 transitions by using a susceptibility motivated by the four-point susceptibility $\chi_4$ studied in glasses \cite{Whitelam2004}, elasto-plastic models \cite{Martens2011, Nicolas2014} and vertex models \cite{Sussman2018}. We define $\tilde{\chi}_4$ as a normalised variance of the number $n_{T1}(\tau)$ of T1 transitions in the time-window $\tau$ \cite{Martens2011, Tyukodi2016}:
\begin{align}
\label{eq:chi4}
\tilde{\chi}_4(\tau)&\equiv \frac{\left\langle (n_{T_1}(\tau) - \langle n_{T_1}(\tau)\rangle)^2\right\rangle}{\langle n_{T_1}(\tau)\rangle}  \quad .
\end{align}
If T1 transitions were completely independent, the variance would be equal to the mean at any $\tau$ and $\tilde{\chi}_4(\tau)= 1$. For T1 transitions organised in avalanches $\tilde{\chi}_4$ grows and reaches a maximum at $\tau_A$, corresponding to the typical avalanche duration. The value at the maximum can be interpreted as a characteristic avalanche size $S \equiv \tilde{\chi}_4(\tau_A)$. After an avalanche, the stress has relaxed locally and new avalanches are  less likely to occur. This effect  leads to a decay of $\tilde{\chi}_4(\tau)$ for $\tau > \tau_A$.

\begin{figure}[ht!]
\centering
\includegraphics[width=.48\textwidth]{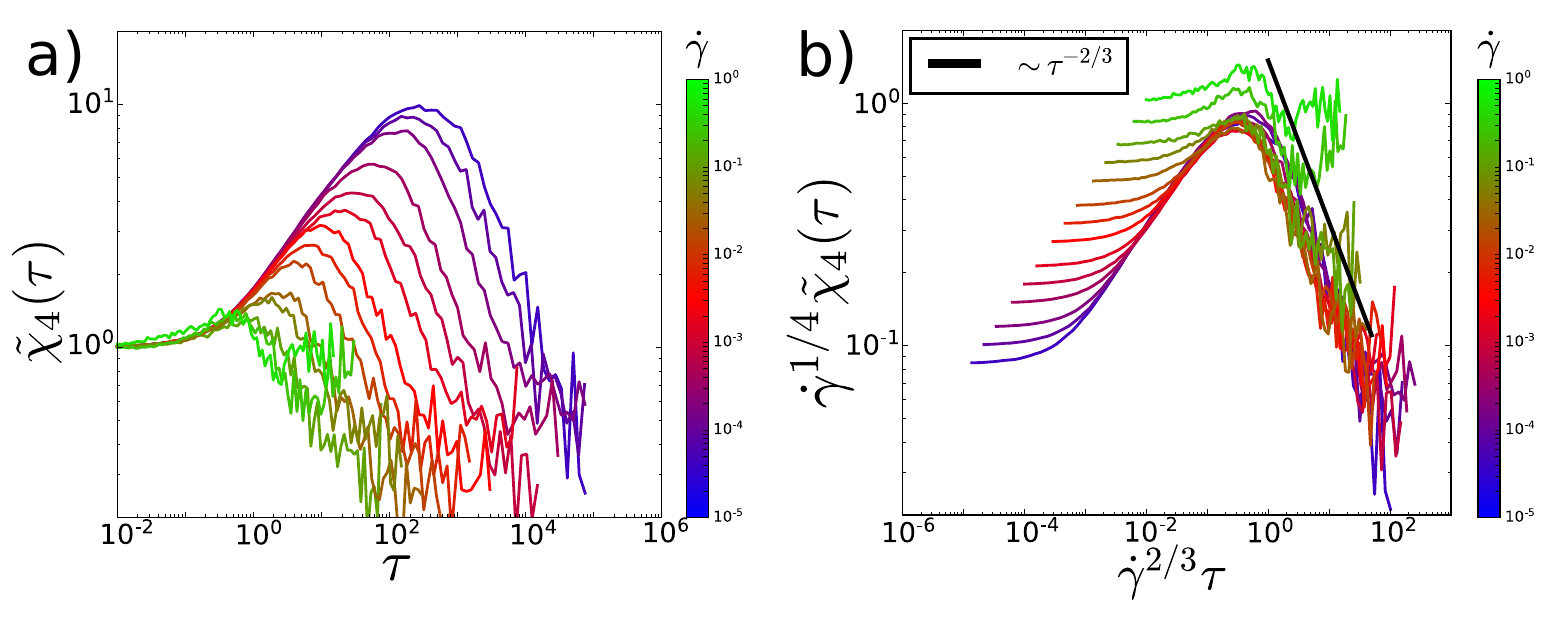}
\caption{ \textbf{a)} The susceptibility $\tilde{\chi}_4(\tau)$ in the steady state flow at different strain rates grows with $\tau$ to a maximal value, which can be interpreted as the avalanche size $S$ of T1 transitions. \textbf{b)} The collapse of $\tilde{\chi}_4(\tau)$ curves after rescaling axes shows that the characteristic avalanche size diverges as $S\sim \dot{\gamma}^{-1/4}$ and that the characteristic avalanche duration diverges as $\tau_A \sim \dot{\gamma}^{-2/3}$. Measurements shown are obtained with $N= 400$.}
\label{fig:exp}
\end{figure}

We measure $\tilde{\chi}_4(\tau)$ in steady state shear flow at different strain rates, see Fig \ref{fig:exp} a) and b).
We find that $\tilde{\chi}_4$ displays a peak which grows near the transition point $\dot\gamma\rightarrow 0$, indeed supporting the idea that the dynamics becomes collective at that point.
Furthermore, we find that the $\tilde{\chi}_4$ curves at different strain rates  collapse when re-scaling the axes as $\dot{\gamma}^{2/3} \tau$ and $\dot{\gamma}^{1/4} \tilde{\chi}_4$. Therefore, as the strain rate  vanishes, the mean avalanche size diverges as $S \sim \dot{\gamma}^{-1/4}$ and the mean avalanche duration as $\tau_A\sim \dot{\gamma}^{-2/3}$.

\section*{Fly wing epithelia}

We have shown that in the vertex model, the bond length distribution $P(L)$ is indicative of the regime in which the material flows:
it presents a singular distribution approaching the solid phase, where the dynamics becomes collective and the flow curve is non-linear. 

\begin{figure}[ht!]
\centering
\includegraphics[width=.48\textwidth]{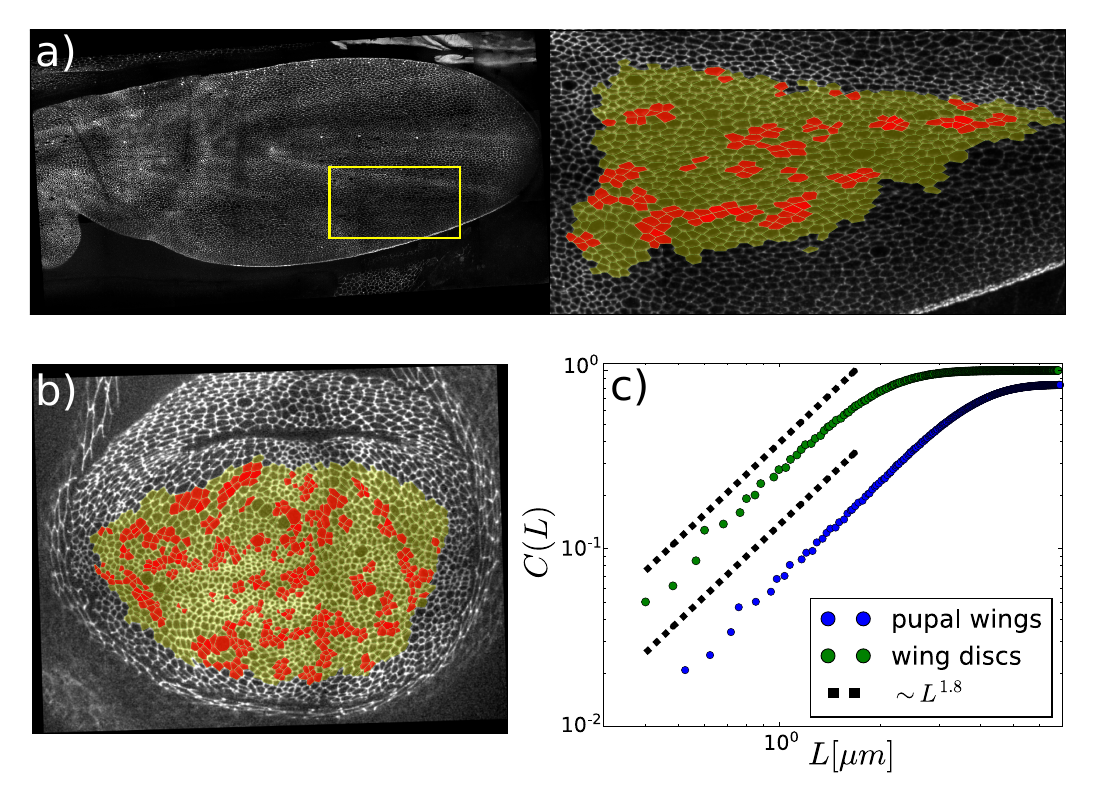}
\caption{\textbf{a)} Pupal wing and \textbf{b)} wing disc of a fruit fly with regions used for analysis highlighted in yellow. Red colored cells have either lost or gained a bond as a part of T1 transition in the last $5$ minutes. \textbf{c)} Cumulative bond length distribution in developing fruit fly wing at the pupal stage of development (blue) and in the larval wing disc (green) both show a clear power-law scaling of $P(L)$ at small bond lengths. Measured effective is in the range $\theta^{\text{eff.}} \approx 0.7-0.9$.}
\label{fig:flow}
\end{figure}

As a first test of the relevance of these ideas to real tissues, we analyse the bond length distribution in wing epithelium of the fruit fly at two stages of development: \textit{i)} during pupal wing morphogenesis, imaged \textit{in vivo} \cite{Etournay2015} and ii) the wing disc epithelium, in third instar larva wing disc epithelium imaged \textit{ex vivo}  \cite{Dye2017}. In the pupal wing we considered a region defined by the longitudinal veins denoted L4 and L5, and the posterior crossvein (yellow cells in Fig. \ref{fig:flow} a), imaged at 5 min intervals between 19 and 23 hours after puparium formation, collected from 3 experiments \cite{Etournay2016}. 
In Fig. \ref{fig:flow} b) we show the analysed region in the wing disc epithelium, corresponding to the wing disc pouch, imaged at 5 min intervals over about 13 hours and collected from 5 experiments. In both Figs. \ref{fig:flow} a) and b) we indicate in red the cells that have lost or gained a bond as a part of T1 transition in the last $5$ min (the time resolution of experiments).

Note that the wing disc epithelia have been developing for about 100 hours before the imaging has started \cite{Dye2017}, while pupal wings have undergone a significant three-dimensional shape change, called eversion \cite{Waddington1940}, before the pupal morphogenesis. Therefore, the initial state of these tissues at the beginning of the experiments could already contain a significant strain history. Thus, although the amount of strain accumulated during experiments is small ($\sim 0.1-0.2$ for pupal wings \cite{Etournay2015, Etournay2016} and $\sim 0.05-0.1$ for wing discs \cite{Dye2017}), it is not clear if we are in a large or small strain regime, with respect to the strain where Fig. \ref{fig:stability} c) displays a minimum. 

Remarkably, we find that in both tissues the bond length distribution $P(L)$ vanishes at small $L$ with the effective exponent $\theta^{\text{eff.}}\approx 0.7-0.9$ (see Fig. \ref{fig:flow} c) and SI for details), similar to those measured in the slowly flowing vertex model at small or large strain. 
This observation suggests that the scaling relation $x\sim L$  holds in real tissues as well, in the range of $L$ that we can probe. Clearly for very small bonds, this relationship must eventually break down due to finite size of vertices \cite{Furuse2014, Bosveld2018}.

It would be very interesting to test this scaling relation directly by perturbing the system. It could be achieved by observing tissue response to a localised mechanical perturbation, such as laser ablation of a single bond: if the energy landscape were smooth with no cusps, the stiffness of the corresponding displacement mode would vanish in approach to a T1 transition as expected near a saddle node bifurcation. As a consequence, a strong locally heterogeneous response would be observed in the experiment at locations of short bonds just before they rearrange, as observed in particulate amorphous solids \cite{Maloney06a} preceding a plastic event. On the other hand, in presence of a cusp there would be no softening and no strong locally heterogeneous displacement preceding cell rearrangements.

Our observation also raises the possibility that developing tissues can lie in a non-linear regime with $\delta \dot{\gamma}/\delta \Sigma \gg \dot{\gamma}/ \Sigma$,  where collective effects are important. Unfortunately, such collective effects are very hard to measure in our experimental data, because the strain rate is not stationary, leading to difficulties using the definition of $\tilde{\chi}_4$. Thus, it would be important to look for non-linear effects more directly by measuring stress dynamics using laser ablation experiments and comparing them to elastic and plastic flow components \cite{Etournay2015, Merkel2017, Blanchard2009, Guirao2015}. Alternatively, epithelia obtained as cultured cell monolayers might provide interesting experimental systems that allow for direct rheological experiments. In such systems non-linear flow properties have been observed \cite{Harris2012} and  bond length distributions $P(L)$ could be measured in different flow regimes.

\section*{Discussion}
We have shown that in the vertex model in its solid phase, the distribution  of bond lengths provides the distribution of local distances to yield stress $x$.
The distribution $P(x)$ reveals properties of the regime in which flow is occurring. This result has two consequences.

First in developing epithelia, we observe a singular distribution of bond lengths, consistent with the one found in the vertex model.
This result  raises the intriguing possibility that non-linear collective effects may be important in tissues, and suggests further empirical tests. Yet, to  precisely relate geometry to flow properties would require us to incorporate active forces and cell divisions or extrusions \cite{Ranft2010, Matoz-Fernandez2017}. From a theoretical perspective, how the density of weak spots depends on stress in the presence of noise - even a simple thermal noise - is not well understood in amorphous solids, and is just starting to be investigated \cite{Chattoraj2010, Matoz-Fernandez2017}. In this light, it would be important to study in the future how different kinds of noise affect the flow curve and the distribution $P(L)$ in the vertex model.

Secondly, despite the fact that the energy functional of the vertex model is  more complex than that of usual particulate materials (in which interaction can be radially symmetric), the relationship between geometry and the presence of weak spots is much simpler in the  vertex model. Outstanding questions in the context of amorphous materials, such as predicting how amorphous solids break by forming shear bands in which most plastic events occur, are hampered by the difficulty of measuring the distribution of weak spot with small $x$ \cite{Cubuk15,Gartner16,Patinet16,Wijtmans17,Schwartzman-Nowik2019}. The vertex model may thus be ideal to understand the universal aspects by which  amorphous materials break and flow.

\begin{acknowledgments}
We thank Tom de Geus, Elisabeth Agoritsas, Wencheng Ji, Matthias Merkel and Ezequiel Ferrero for useful comments and discussions. M.W. thanks the Swiss National Science Foundation for support under Grant No. 200021-165509 and the Simons Foundation Grant No. 454953. N.A.D. acknowledges funding from the Deutsche Forschungsgemeinschaft (EA4/10-1, EA4/10-2).
  \end{acknowledgments}

\bibliography{bib,Wyartbibnew}
\newpage
\appendix

\onecolumngrid
  \vspace{1cm}
  \begin{center}
    \Large
    \textbf{Supplementary Information}
  \end{center}
  \vspace{1cm}
\twocolumngrid
  
\subsection{Vertex model parameters}
We used the following parameter values in most simulations:
\begin{itemize}
\item $K_c= 10$ 
\item $\Lambda_b= -11$
\item $A_{0, c}$ and $\Gamma_c$ were always chosen so that $p_0= -\Lambda_b/(2 \Gamma_c \sqrt{A_{0, c}})= 11/(2 \sqrt{3 \sqrt{3}/2})\approx 3.41$ is constant throughout the network.
  \item $\nu= 1$
  \end{itemize}
Any change of parameters in a particular simulation is explicitly listed below.
  
\subsection{T1 transition implementation}
When a bond length becomes smaller than a threshold value $\epsilon_{T_1}$ a T1 transitions is attempted: old bond and corresponding vertices are destroyed and  new ones are created, then forces on the new bond are computed and T1 transition is allowed if the tension in the new bond is positive (forces are stretching the new). Otherwise, the T1 transition is canceled and the network is reverted to the original state.
The choice of $\epsilon_{T1}$  is specified for particular simulations below. To avoid the possibility of an extrusion we do not allow bond loss by T1 transition for cells with 3 neighbors.

\subsection{Cell size polydispersity}
In the flow simulations we avoided crystallization and shear banding by introducing cell size polydispersity: preferred cell areas $A_{0,c}$ are uniformly distributed on the interval $[\bar{A}_{0,c}/2, 3 \bar{A}_{0, c}/2]$, where the mean preferred cell area $\bar{A}_{0, c}= 3 \sqrt{3}/2$ corresponds to size of regular hexagons before initial network randomization. Parameter $\Gamma_c$ was then chosen so that all cells have the same value of $p_0$. Finally, we set $K_c= 40$.

\subsection{Isotropic disordered networks}
To create isotropic disordered networks, we first create a hexagonal network with bond length $l=1$. The energy function parameters are then set so that the network is in the solid phase $p_0 \approx 3.7$, but close to the transition point $p_0^{*} = 3.81$ \cite{Bi2015}. 
The parameters used are:
\begin{itemize}
	\item $K_c = 1$ 
	\item $\Lambda_b = -12$ 
	\item $\Delta t= 5 \cdot 10^{-4}$
	\item $\epsilon_{T1} = 5 \cdot 10^{-4}$ 
\end{itemize}

For the randomization process we introduce fluctuations of the bond tension $\Lambda$ independently in each bond by simulating its dynamics as a time-discretised Ornstein-Uhlenbeck process:
\begin{align}
\label{eq:1}
\Lambda(t + \delta t)&= \Lambda(t) - k(\Lambda (t) - \Lambda_0) + \xi \sqrt{\delta t} ,
\end{align}
where $\Lambda_0= -12$, $\delta t= 10^{-2}$, $k= 1$ and $\xi$ a random variable taken from a normal distribution $\mathcal{N}(0, 1)$.

Networks are evolved for $t_r= 50$, then fluctuations are frozen and the network is relaxed for time $0.5$. Finally, network parameters are set to simulation values and networks are further relaxed over time using $\epsilon_{T1}= 10^{-4}$ until the net force on any vertex was below $\epsilon_F= 10^{-4}$. 

\subsection{Stress redistribution by a T1 transition}

Results of the stress redistribution from a T1 event were obtained from a $N= 2500$ network. After initial relaxation (until net force on any vertex is below $\epsilon_F = 10^{-5}$) 50 bonds of length $l<0.1$ were selected randomly. Each of these bonds was shrunk below the T1 threshold. If the tension in the newly formed bond was negative (so that T1 transition would revert back, see Section A) the bond was not considered, otherwise the T1 was performed and the network was relaxed (with $\epsilon_F= 10^{-4}$) and shear stress change for each cell was recorded. We defined cell position as the position of its center of mass and the position was recorded in a coordinate system centered at the middle of the original bond that was selected and whose $x$-axis was aligned with the original bond direction. Finally, we obtained the average shear stress change in space by performing a cell area-weighted average in spatial bins.  

\subsection{Direct measurement of P(x)}
In a subset of isotropic disordered networks each bond of length $L<0.5$ was selected and constrained to be of length $L_c=10^{-6}$. With this constraint on the bond, the network is then relaxed (until the net force on any vertex was below $\epsilon_F= 10^{-3}$).
Once the relaxation is over, the magnitude of the force acting on the vertices of the constrained bond is recorded as $x$, as well as network energy change $\Delta E$.

\subsection{Steady state shear flow}
We apply simple shear strain to a network at each time-step using an affine transformation of all vertex positions:
\begin{align}
\label{eq:2}
\Delta x_{\alpha}&= \frac{y_{\alpha}}{L_y}\Delta \gamma ,
\end{align}
where $(x_{\alpha}, y_{\alpha})$ are coordinates of vertex $\alpha$, $\Delta\gamma$ is strain increment, and $L_y$ simulation box size in $y$ direction. The applied strain rate $\dot{\gamma}=\Delta \gamma / \Delta t$ was always constant during a simulation. T1 transition threshold was $\epsilon_{T1}= 10^{-2}$ and simulation time step $\Delta t= 10^{-3}$.

\subsection{Isotropic $\theta$}
We determined the range of values corresponding to this exponent by fitting cumulative bond length distribution obtained from 50 isotropic networks of size $N= 2500$, which exhibit a broader range of scaling than $N= 400$ networks. We performed the power law fit on a range of data $[l, 0.2]$ with varying lower limit $l$ as shown in Fig. \ref{fig:expDataFits} a). We find that for values of lower limit $l$ in the range $[0.01, 0.05]$ the exponent $\theta$ varies between $0.5$ and $0.60$.

\subsection{Transient $\theta_L^{\text{eff.}}$}
We fitted the effective exponent on cumulative distribution of adjusted bond lengths $L_r= L - \epsilon_{T1}$ in the range $[0.03, 0.3]$, accumulated from 700 realisations obtained in $N= 400$ networks at strain rate $\dot{\gamma}= 10^{-4}$, recorded at strain resolution $\delta\gamma_1= 0.02$. The plot in Fig. 3 c) was obtained by averaging the results in windows of width $\delta\gamma_2=0.1$ with the shaded regions indicating the corresponding standard deviation in each window.

\subsection{Steady state $\theta$}
Cumulative bond length distribution of adjusted bond lengths $L_r= L - \epsilon_{T1}$ shown in Fig. 3 d) of the main text are obtained from 1400 networks at strain beyond 5 taken at strain intervals $\Delta \gamma= 0.02$. For the lowest strain rate $\dot{\gamma}= 5\cdot 10^{-5}$ we fitted a power law on a range of data $[l, 0.3]$ with varying lower limit $l$ as shown in Fig. \ref{fig:expDataFits} b). We find that for values of lower limit $l$ in the range $[0.02, 0.05]$ the exponent $\theta$ varies between $0.56$ and $0.77$.

\subsection{Effective exponent in experimental data}
\begin{figure}
\centering
\includegraphics[width=.7\textwidth]{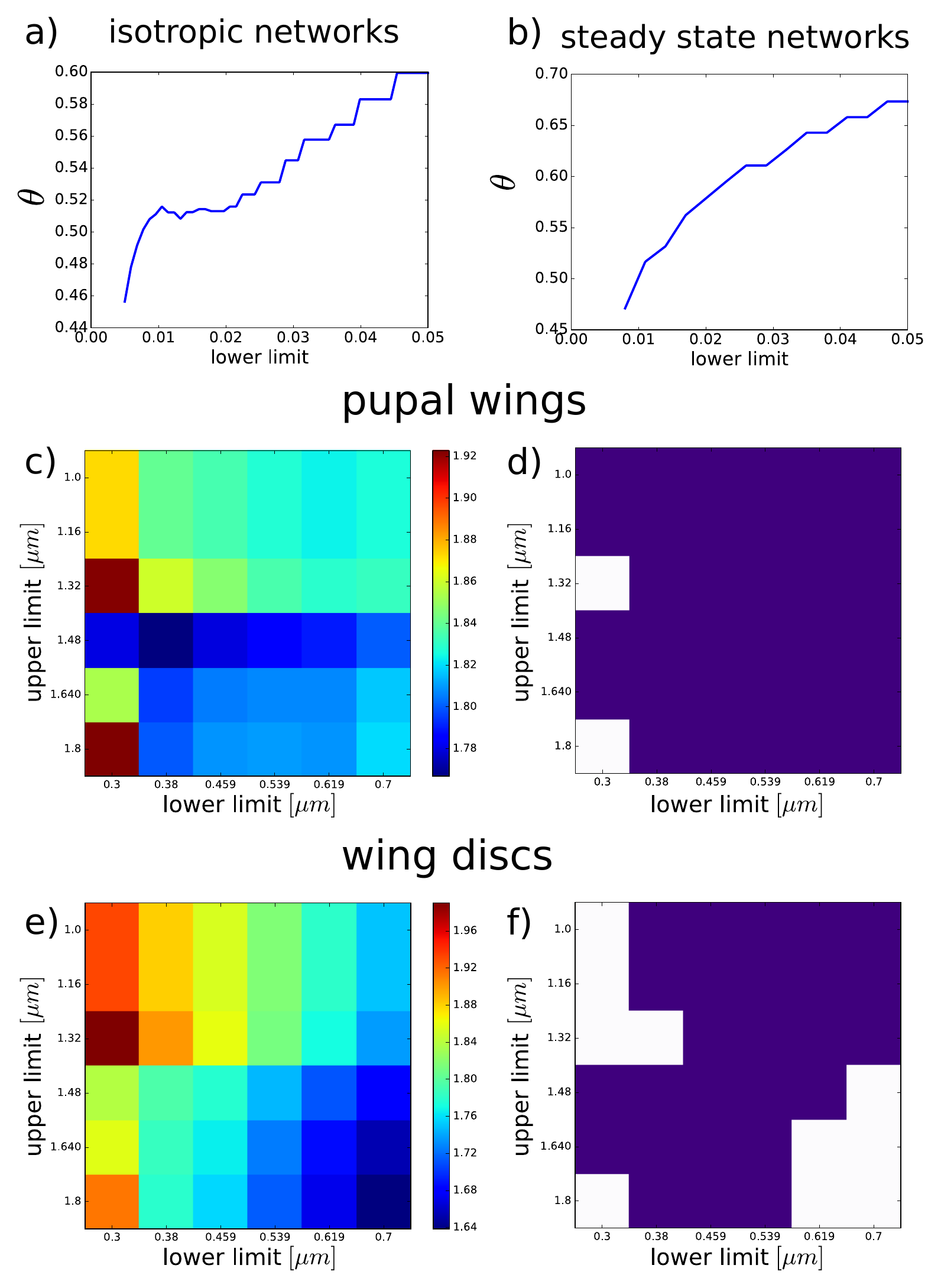}
\caption{\textbf{a)} and \textbf{b)} Values of exponent $\theta$ obtained by fitting cumulative bond length distribution in isotropic networks and steady state networks, respectively,  using different values of lower limit $l$ of data range (see SI text above).  Values of the effective exponent $\theta^{\text{eff.}}$ obtained by fitting the cumulative distribution of bond lengths in experimental data. Subplots \textbf{c)} and \textbf{e)} show the values of the exponent in pupal wigs and wing discs, respectively. Subplots \textbf{d)} and \textbf{f)} show in purple choices of data limits for which the exponent value falls in the range $0.7 - 0.9$ in pupal wigs and wing discs, respectively.}
\label{fig:expDataFits}
\end{figure}

To determine the effective exponent $\theta_L^{\text{eff.}}$ in experimental data we fitted a power law to the cumulative distribution of bond lengths. We vary the fitting range to estimate confidence interval of the fits and we find that in most cases fitted exponents fall in the range $0. 7 < \theta_L^{\text{eff.}} < 0.9$, as shown in Fig. \ref{fig:expDataFits} c) - f).
\onecolumngrid

\end{document}